\documentclass[aip,jcp,reprint,floatfix]{revtex4-1}
\usepackage{amsmath,amssymb}
\usepackage{epsfig}
\usepackage{graphicx}
\usepackage{dcolumn}
\usepackage{bbold}
\usepackage{natbib}
\usepackage[colorlinks=true,citecolor={blue},urlcolor={blue}, linkcolor=blue]{hyperref}%

\begin{document}
\title{Relativistic equation-of-motion coupled-cluster method using open-shell reference wavefunction}%
%%%%%%%%%%%%%%%%%%%%%%%%%%%%%%%%%%%%%%%%%%%%%%%%%%%%%%%%%%%%%%%%%%%%%%%%%%%%%%%%%%%%%%%%%%%%%%%%%%%%%%%%%%%%%%%%%%%%%%%%%%%%%%%%%%%%%%
\author{Himadri Pathak}%
\thanks{hmdrpthk@gmail.com}
\affiliation{Electronic Structure Theory Group, Physical Chemistry Division, CSIR-National Chemical Laboratory, Pune, 411\,008, India}%
%%%%%%%%%%%%%%%%%%%%%%%%%%%%%%%%%%%%%%%%%%%%%%%%%%%%%%%%%%%%%%%%%%%%%%%%%%%%%%%%%%%%%%%%%%%%%%%%%%%%%%%%%%%%%%%%%%%%%%%%%%%%%%%%%%%%%%%
\author{Sudip Sasmal}
\thanks{sudipsasmal.chem@gmail.com}
\affiliation{Electronic Structure Theory Group, Physical Chemistry Division, CSIR-National Chemical Laboratory, Pune, 411\,008, India}
%%%%%%%%%%%%%%%%%%%%%%%%%%%%%%%%%%%%%%%%%%%%%%%%%%%%%%%%%%%%%%%%%%%%%%%%%%%%%%%%%%%%%%%%%%%%%%%%%%%%%%%%%%%%%%%%%%%%%%%%%%%%%%%%%%%%%%%
\author{Malaya K. Nayak}
\affiliation{Theoretical Chemistry Section, Bhabha Atomic Research Centre, Trombay, Mumbai 400\,085, India}
%%%%%%%%%%%%%%%%%%%%%%%%%%%%%%%%%%%%%%%%%%%%%%%%%%%%%%%%%%%%%%%%%%%%%%%%%%%%%%%%%%%%%%%%%%%%%%%%%%%%%%%%%%%%%%%%%%%%%%%%%%%%%%%%%%%%%%%
\author{Nayana Vaval}
\affiliation{Electronic Structure Theory Group, Physical Chemistry Division, CSIR-National Chemical Laboratory, Pune, 411\,008, India}
%%%%%%%%%%%%%%%%%%%%%%%%%%%%%%%%%%%%%%%%%%%%%%%%%%%%%%%%%%%%%%%%%%%%%%%%%%%%%%%%%%%%%%%%%%%%%%%%%%%%%%%%%%%%%%%%%%%%%%%%%%%%%%%%%%%%%%%%
\author{Sourav Pal}
\thanks{spal@chem.iitb.ac.in}
\affiliation{Department of Chemistry, Indian Institute of Technology Bombay, Powai, Mumbai 400\,076, India}
\begin{abstract}
The open-shell reference relativistic equation-of-motion coupled-cluster method within its
four-component description is successfully implemented with the consideration of single- and double- excitation approximation.
The one-body and two-body matrix elements required for the correlation calculation are generated using Dirac-Coulomb Hamiltonian. 
As a first attempt, the implemented method is employed to calculate a few of the low-lying ionized states of heavy
atomic (Ag, Cs, Au, Fr, Lr) and valence ionization potential of molecular (HgH, PbF) systems,
where the effect of relativity does really matter to obtain highly accurate results.
Not only the relativistic effect, but also the effect of electron correlation is crucial in these heavy atomic and molecular systems. To justify the
fact, we have taken two further approximations in the four-component relativistic equation-of-motion framework to quantify how the
effect of electron correlation plays a role in the calculated values at different level of the approximation.
All these calculated results are compared with the available experimental data as well as with other theoretically calculated values
to judge the extent of accuracy obtained in our calculations.
A nice agreement is achieved for all the considered systems which validates the performance of the open-shell reference
relativistic equation-of-motion coupled-cluster method.
%The outcome of our intermediate calculation in the equation-of-motion framework is clearly suggests that the non-dynamic part of the electron correlation
%has a larger role than the dynamic part in the computed ionization potential values of the considered heavier systems.
%%%%%%%%%%%%%%%%%%%%%%%%%%%%%%%%%%%%%%%%%%%%%%%%%%%%%%%%%%%%%%%%%%%%%%%%%%%%%%%%%%%%%%%%%%%%%%%%%%%%%%%%%%%%%%%%%%%%%%%%%%%%%%%%%%%%%%%%%%
%\pacs{31.15.bw, 31.15.vn, 33.15.Ry}
%%%%%%%%%%%%%%%%%%%%%%%%%%%%%%%%%%%%%%%%%%%%%%%%%%%%%%%%%%%%%%%%%%%%%%%%%%%%%%%%%%%%%%%%%%%%%%%%%%%%%%%%%%%%%%%%%%%%%%%%%%%%%%%%%%%%%%%%%%
\end{abstract}
%\date{\today}
\maketitle
\section{Introduction}
The accurate calculations of the transition energies, especially for the heavier atomic and molecular systems require
simultaneous incorporation of  both the effect of relativity and electron correlation as there is a sinuous ally between these
two effects \cite{1}. The Dirac-Hartree-Fock (DHF) method is the best suited recipe to include the effect of relativity in a single 
determinant wavefunction based theory using four-component Dirac spinor. In the DHF technique, the complicated many-electron problem
is converted into sum of many one-electron problems and an average interaction between the electrons is assumed in the due course.
Therefore, the instantaneous interaction between the electrons having opposite spin or the electron correlation is missing in such a treatment.  
On the other hand, equation-of-motion coupled-cluster (EOMCC) method is very successful in the non-relativistic theory for the treatment
of electron correlation. The performance of the method is well tested for the  calculations of ionization potential, electron affinity and excitation energy
starting from both closed-shell as well as open-shell reference wavefunction \cite{2,3,4,5,6,7,8,9,10,11,12,12a,12b}
but the full fledged extension of the EOMCC method
to the relativistic realm is far fewer for the former \cite{13,14,15} and none for the latter. \par
%%%%%%%%%%%%%%%%%%%%%%%%%%%%%%%%%%%%%%%%%%%%%%%%%%%%%%%%%%%%%%%%%%%%%%%%%%%%%%%%%%%%%%%%%%%%%%%%%%%%%%%%%%%%%%%%%%
The EOMCC method is a very handy technique since the difference of energy is calculated in a direct manner.
A two step procedure is followed in actual practice to solve the EOMCC amplitude equations as ground state
amplitudes are decoupled from that of the excited states;
(i) solution of the $N$ electron reference wavefunction and (ii) 
the diagonalization of the constructed effective Hamiltonian matrix in the ($N-1$) electron
determinantal space for the single electron ionization problem.
%Therefore, the method is taking account of two Hilbert space simultaneously, which fortifies the inclusion 
%of relaxation effect.
The idea of diagonalization of the effective Hamiltonian matrix elements 
is, by and large, associated with the multi-reference theories, however, the EOMCC method works within a single reference 
description to tackle complex wavefunction which is multi-configurational in nature.
Furthermore, the effect of relaxation which has a definite role in assigning the ionized states precisely is also taken care by the EOMCC method.
%Relaxation effect have definite role in assigning the ionized states precisely
However, it has a greater relevance in ionization of an inner orbital electron as compared to the valence electrons.
In principle, the secondary ionized states or the excitation energies can also be obtained as a by-product,
which have key role in explaining various photo-ionization spectra. 
%Alternatively, we may say that the EOMCC method for the single electron ionization problem
%is a transition of an electron from the occupied orbitals to continuum.
It is worth to note that EOMCC is a size consistent method but the calculated transition energy 
is extensive only for the principal valence sector whereas the secondary roots are size intensive \cite{16,17,18}. 
We can say that the EOMCC is 
a kind of configuration interaction (CI) method using the coupled-cluster similarity transformed Hamiltonian 
matrix elements. Therefore, the convergence difficulty is not of concern due to structural resembles with the CI method, which
is always an integral part of any multi-root theories. However, there are few other alternative choices to deal with such a problem, a brief overview of various
options can be found in Ref. \cite{18a}. \par

%%%%%%%%%%%%%%%%%%%%%%%%%%%%%%%%%%%%%%%%%%%%%%%%%%%%%%%%%%%%%%%%%%%%%%%%%%%%%%%%%%%%%%%%%%%%%%%%%%%%%%%%%%%%%%%%%%%%%%%%%%%%%%%%%%%%%%%%%%%%
The Fock-space multi-reference coupled-cluster method (FSMRCC) \cite{19,20,21,22,23,24} is an alternative approach to calculate transition energies 
between an arbitrary chosen ground state and several other final states. Operationally, this approch is divided 
into various ``sectors'' depending on the number of electron added to or removed from the system.
In the FSMRCC approach, the calculation of a given sector (x,y) depends upon all (a,b); $a\le x$ and  $b\le y$, sector amplitude equations.
Therefore, this advocates for a hierarchical solution. 
On the other hand, the EOMCC method deals only with the amplitudes of the reference sector and that of the target Fock space sector of interest.    
Therefore, these two approach eventually produces identical results for the single electron ionization or attachment problem but different 
for the excitation of electron from an occupied orbital to a virtual orbital and in multi-electron ionization or attachment processes.
It appears that FSMRCC theory is conceptually difficult in contrast to the EOMCC method which is very easy to understand.
Therefore, it can be used in a ``black box'' way to address various electronic structure problems  
which motivates us to focus in the EOMCC domain. 
A detail comparative discussion in the context of performance between these two methods
can be found in  Ref. \cite{25}.
Stanton and co-workers \cite{26} implemented FSMRCC theory for general single determinant reference wavefunction and applied 
to calculate ionization potential values of molecules starting from high spin ground state configuration but their approach lacks the consideration of 
relativistic motion of the electrons. On the other hand, Kaldor and co-workers \cite{27,28,29,30} have done extensive work in the relativistic 
FSMRCC framework, but they always starts from a closed-shell reference wavefunction. As for example, they applied (1,0) sector 
FSMRCC theory to a mono-cationic closed-shell reference state and negative of the obtained value is reported 
as the ionization potential value of the open-shell atomic or molecular systems. We have also followed such an idea in a recent work using EOMCC method \cite{30a}. 
The problem with such an approach is that it can only be applied for the calculation of valence ionization potential of the open-shell atomic and molecular 
system but the ionization of inner orbital electrons can not be addressed, those states are more interesting due to the presence of strong spin-orbit coupling (SOC).
Moreover, either (1,0) sector FSMRCC theory or the electron attached (EA) EOMCC method require
four particle intermediate H$_{\text {eff}}$ matrix elements. The size of those matrix element and the required computational time due to
its scaling is enormous particularly for a relativistic calculation. Therefore, it is not at
all a good idea to follow such an approach from the computational perspective, especially in the relativistic domain.
On the other hand, starting from an open-shell reference wavefunction is a more
realistic choice, particularly when large non-dynamic electron correlation effects are present. \par
%%%%%%%%%%%%%%%%%%%%%%%%%%%%%%%%%%%%%%%%%%%%%%%%%%%%%%%%%%%%%%%%%%%%%%%%%%%%%%%%%%%%%%%%%%%%%%%%%%%%%%%%%%%%%%%%%%%%%%%%%
The most common practice in a molecular relativistic calculations is the use of relativistic effective core potential (RECP) with
the inclusion SOC \cite{31,32}. The SOC is the dominant part of the relativistic effect and play a significant role specially 
for the heavy p-block elements and also for the moderate  heavy molecular systems when very accurate results are soughted.
Due to the effect of SOC, different degenerate electronic configuration mixed with each other and leads to significant change in 
the wavefunction, energetics and spectra. 
Hirata {\it et al,} \cite{33} made the first attempt to include SOC with RECP in the EOMCC framework.
They made a composite method by taking consideration of different electron correlation 
methods, basis set, and a part of the relativistic effect is supplemented.
The ground state wavefunction is constructed at the second-order perturbation theory level
to take care of the dynamic part of the electron correlation with the addition of scalar 
relativistic effect and extrapolated it to the complete basis set limit.
The EOMCC calculation is done to take care of the non-dynamic part of the electron correlation using a very small basis set.
The SOC effect is quantified as the energy difference between RECP+SOC with RECP using the low rank method employed 
to take care of the dynamic part. The applied strategy cancels some of the errors associated with the RECP method.
The RECP is generally build upon the separation of the electrons according to the energy with classification of electron 
into core and valence. The valence including the few outer-core electrons are treated explicitly and a effective RECP
operator is defined to replace the other electrons. This RECP operator takes care of the interaction between the
explicitly treated electrons with the other excluded electrons.
Therefore, a large number of less contributing electrons are excluded from the calculations and reduces 
the computational cost as compared to all electron relativistic treatment. 
However, the approach can not guarantee high accurate results as all the electrons are not treated explicitly.
Therefore, neither the effect of electron correlation nor the relativistic effect is comprehensive in their approach. 
In a recent work Epifanovsky {\it et al.} \cite{34} did perturbative treatment to include SOC with explicit treatment of all the electrons
using Breit-Pauli Hamiltonian with zeroth-order non-relativistic wavefunction in different variant of EOMCC method and applied to
a few very small molecular systems where the effect of relativity does not have much role to play.
Li {\it et al.} \cite{35} included SOC only in the EOM part to calculate ionization potential of molecular system but they have taken 
$N$ electron closed-shell coupled-cluster wavefunction as a reference.  
The perturbative treatment of SOC is not reliable for heavy atomic and molecular systems. Therefore, a fully relativistic 
description is imperative to address such effects. The solution based on Dirac-Hamiltonian naturally takes care of the SOC effect \cite{36}. 
The effect is more prominent for the  open-shell systems as compared to the closed-shell atomic or molecular systems which
motivates us to implement EOMCC method starting from an open-shell reference wavefunction
using four-component Dirac spinors with the explicit treatment of all the electrons in the SCF calculations
and to apply for the treatment of heavy atomic and molecular systems. \par
%%%%%%%%%%%%%%%%%%%%%%%%%%%%%%%%%%%%%%%%%%%%%%%%%%%%%%%%%%%%%%%%%%%%%%%%%%%%%%%%%%%%%%%%%%%%%%%%%%%%%%%%%%%%%%%%%%%%%%%%%%%%
In the present article, discussion is only restricted to implementation of relativistic EOMCC method in the context 
of single electron ionization problem starting from an open-shell reference wavefunction, though it can be 
generalized to any other Fock-space sector of interest by defining an appropriate EOM operator, and its 
application to heavy atomic and molecular systems where the effects of relativity have a much decisive role in the precise
calculation of ionization potential.
The reference wavefunction in the EOMCC method is taken as the
single reference coupled-cluster ground wavefunction. We have constructed the coupled-cluster ground state wavefunction
considering up to single- and double- excitation (CCSD) from N electron Dirac-Hartree-Fock (DHF) reference wavefunction.
The EOM matrix elements are constructed in the 1h and 2h-1p space and diagonalized to get a few of the low-lying principal roots. 
Beside the effect of relativity, 
correlation effects also have important role in the heavier systems. We, therefore, followed two intermediate
scheme in the four-component relativistic EOMCC framework to understand the effect of electron correlation in the calculated ionization potential 
values. We call these as EOM-MBPT(2) and EOM-RPA.
In the EOM-MBPT(2) scheme, the ground state wavefunction is constructed using first order perturbed 
wavefunction, which corresponds to the second order perturbation (MBPT(2)) energy as the ground state energy. On the other hand,
in the EOM-RPA scheme, the ground state wavefunction is constructed at the CCSD level but the EOM matrix elements 
are constructed only in the 1h space. Therefore, in the EOM-MBPT(2) scheme, a part of the dynamic correlation is missing whereas
in the EOM-RPA scheme, dominant part of the non-dynamic correlation, since
contribution from the 2h-1p block is absent which contributes significantly to the non-dynamic part of the electron correlation.
Finally, all these results are compared with the values calculated by means of other theoretical methods and with the values from experiment. 
These three calculations will help us to understand how the effect of electron correlation at different level of approximation is equally important for the heavier systems.  
We have taken a few representative example of heavy atomic (Ag, Cs, Au, Fr, Lr) and molecular (HgH, PbF) systems where both relativistic effect and electron correlation
have a larger role in getting ionization potential values accurately. \par
%%%%%%%%%%%%%%%%%%%%%%%%%%%%%%%%%%%%%%%%%%%%%%%%%%%%%%%%%%%%%%%%%%%%%%%%%%%%%%%%%%%%%%%%%%%%%%%%%%%%%%%%%%%%%%%%%%%%%%%%%%%%%%%%%%
This paper is organized as follows. A brief overview 
of the EOMCC method in the context of single electron ionization problem is discussed 
in Sec. \ref{sec2}. Sec. \ref{sec3} and Sec. \ref{sec4} are reserved for details of the 
computational consideration and discussion of the obtained results in our calculations, respectively.
Finally we made our concluding remark in Sec. \ref{sec5}. We are consistent with atomic unit if not
declared. 
\section{Method} \label{sec2}
In the following section we present a brief overview of the EOMCC method in the context of ionization
of a single electron from the $N$ electron coupled-cluster reference wavefunction, and the final algebraic equations
are represented in their spinor form. In the EOMCC method, a linear operator ($R$) acts on the reference
wavefunction to generate the excited state determinants. We have consistently used  $i,j (a,b)$ indices
to represent the occupied (virtual) spinors. \par
%%%%%%%%%%%%%%%%%%%%%%%%%%%%%%%%%%%%%%%%%%%%%%%%%%%%%%%%%%%%%%%%%%%%%%%%%%%%%%%%%%%%%%%%%%%%%%%%%%%%%%%%%%%%%%%%%%%%%%%%%%%%%%%%%%%%%%%%
The Schr\"{o}dinger equation for the reference state and the $\nu^{th}$ excited state can be represented as  
%%%%%%%%%%%%%%%%%%%%%%%%%%%%%%%%%%%%%%%%%%%%%
\begin{eqnarray}
{H}|\Psi_0\rangle=E_0|\Psi_0\rangle,
\label{eqn1}
\end{eqnarray}
%%%%%%%%%%%%%%%%%%%%%%%%%%%%%%%%%%%%%%%%%%%%%%
\begin{eqnarray}
{H}|\Psi_\nu\rangle=E_\nu|\Psi_\nu\rangle,
\label{eqn2}
\end{eqnarray}
%%%%%%%%%%%%%%%%%%%%%%%%%%%%%%%%%%%%%%%%%%%%%%
The excited state wavefunction can be described as 
\begin{eqnarray}
|\Psi_\nu\rangle=R_\nu|\Psi_0\rangle .
\end{eqnarray}
Left multiplication of the Eqn. \ref{eqn1} with $R_\nu$ and subtraction from  Eqn. \ref{eqn2}
will lead to the following expression
%%%%%%%%%%%%%%%%%%%%%%%%%%%%%%%%%%%%%%%%%%%%%%
\begin{eqnarray}
[H,R_\nu]|\Psi_0 \rangle=\Delta E_\nu R_\nu|\Psi_0\rangle .
\label{eom}
\end{eqnarray}
Here, $\Delta E_\nu$ is the energy difference between the reference state and the excited states.
%%%%%%%%%%%%%%%%%%%%%%%%%%%%%%%%%%%%%%%%%%%%%%%%
With proper definition of the $R_\nu$ operator, excited states of electron detached, attached or even excited electronic states can be obtained.
The expression of the $ R_\nu$ operator for the single electron ionization problem is as
\begin{equation}
\begin{split}
R_\nu^{IP} =& R_{1}+R_{2}+\dots\\
             =&  \sum_{i} r_{i} a_i +  \sum\limits_{\stackrel{a}{i < j}} r_{i j}^{a} a^{\dag}_a a_j a_i+\dots .
\end{split}
\end{equation}
%We have approximated upto $R_1$ (1h) and $R_2$ (2h-1p) creation   
%%%%%%%%%%%%%%%%%%%%%%%%%%%%%%%%%%%
The reference wavefunction in the EOMCC method is not necessarily the ground state wavefunction but we 
have chosen the coupled-cluster ground state wavefunction as the reference state. 
The ground state wavefunction in the coupled-cluster method is defined as
\begin{eqnarray}
|\Psi_0\rangle=e^T|\Phi_0\rangle .
\end{eqnarray}
Here, $T$ is the cluster operator and $|\Phi_0\rangle$ is the $N$ electron DHF eigenstate.
The second-quantization representation of the cluster-operator, $T$, is as follows
%The cluster operator $T$ is represented in second quantization form as 
%%%%%%%%%%%%%%%%%%%%%%%%%%%%%%%%%%%%%%%%%%%%%%%%%%%%
\begin{equation}
\begin{split}
T=& T_{1}+T_{2}+\dots\\
             =&  \sum\limits_{{i,a}} t_{i}^{a}a_{a}^{\dag}a_{i} +
     \sum\limits_{\stackrel{a<b}{i<j}}t_{ij}^{ab}a_{a}^{\dag}a_{b}^{\dag}a_{j}a_{i}+\dots .
\end{split}
\end{equation}
The following simultaneous non-linear algebraic equations are iteratively solved to get the cluster amplitudes
%and are graphically presented in Fig \ref{fig3}
\begin{equation}
{\langle \Phi_{i}^{a}|e^{-T} He^{T}|\Phi_{0}\rangle=0,\,\,\, \langle \Phi_{ij}^{ab}|e^{-T} He^{T}|\Phi_{0}\rangle=0 }.
\end{equation}
%%%%%%%%%%%%%%%%%%%%%%%%%%%%%%%%%%%%%%%%%%%%%%%%%%%%%%%%%%
where, $|\Phi_{i}^{a}\rangle$ and $|\Phi_{ij}^{ab}\rangle$ are the singly and doubly excited determinant with reference 
to the DHF determinant.
Finally, the ground state energy is obtained by solving the %${E_{ccsd}=\langle \Phi_{0}|e^{-T}He^{T}|\Phi_{0}\rangle}$
equation for the energy,
\begin{equation}
{E_{ccsd}=\langle \Phi_{0}|e^{-T} He^{T}|\Phi_{0}\rangle} , 
\end{equation}
where, $\hat H$ is the Dirac-Coulomb Hamiltonian which is, 
\begin{eqnarray}
{H_{DC}} &=&\sum_{A}\sum_{i} [c (\vec {\alpha}\cdot \vec {p})_i + (\beta -{\mathbb{1}_4}) m_0c^{2} + V_{iA}] \nonumber\\
       &+& \sum_{i>j} \frac{1}{r_{ij}} {\mathbb{1}_4},%+\sum_{A,B>A} V_{AB},
\end{eqnarray}
here, $\alpha$ and $\beta$ are the known Dirac matrices. $V_{iA}$ is the potential energy operator
for the $i^{th}$ electron in the field of nucleus A.
$m_0c^2$ is the rest mass energy of the free electron, where c stands for the speed of light. \par
%%%%%%%%%%%%%%%%%%%%%%%%%%%%%%%%%%%%%%%%%%%%%%%%%%%%%%%%%%%%%%%%%%%%%%%%%%%%%%%%%%%%%%%%%%%%%%%%%%%%%%%%%%%%%%%%%%%%%%%%%%%%%%%%%%%%%%%%%
The $R_\nu$ and $T$ operator commutes among themselves since they are constructed of same quasi-particle creation operator. Therefore,
we can represent Eqn. \ref{eom} as 
%%%%%%%%%%%%%%%%%%%%%%%%%%%%%%%%%%%%%%%%%%%%%%%%%%%%%%%%%
\begin{eqnarray}
[{\bar H},R_\nu]|\Phi_0 \rangle=\Delta E_\nu R_\nu|\Phi_0\rangle \,\,\, \forall \nu,
\end{eqnarray}
Here, $\bar {H}=e^{-T}He^{T}$ is the effective Hamiltonian. % and $\Delta E_\nu$ is the corresponding ionization 
%potential value.
%%%%%%%%%%%%%%%%%%%%%%%%%%%%%%%%%%%%%%%%%%%%%%%%%%%%%%%%%%%%%
The above equation is projected onto the set of excited determinants ( ($|\Phi_i\rangle$) and ($|\phi_{ij}^{a}\rangle$) 
%A correlated determinantal space of 1h ($|\Phi_i\rangle$) and 2h-1p ($|\phi_{ij}^{a}\rangle$) is chosen
to get the desired ionization potential values ($\Delta E_\nu$).
\begin{equation}
{\langle \phi_{i}|[\bar{H}, R_\nu]|\phi_{0}\rangle=\Delta E_\nu R_i},
%{\langle \phi_{ij}^{a}[\bar {H}R(k)]|\phi_{0} \rangle =\Delta E_k R_{ij}^{a}},
\label{r1}
\end{equation}
\begin{equation}
{\langle \phi_{ij}^{a}|[\bar {H},R_\nu]|\phi_{0} \rangle =\Delta E_\nu R_{ij}^{a}},
\label{r2}
\end{equation}
%%%%%%%%%%%%%%%%%%%%%%%%%%%%%%
The matrix form of the above equations is as
$\bar {H}R = R\Delta E_\nu$.
%The effective Hamiltonian matrix is constructed
%in 1h and 2h-1p space and diagonalized only in
%the 1h space.
The algebraic expression of the left hand sides of Eqn. \ref{r1} and \ref{r2} are as follows  
\begin{widetext}
\begin{eqnarray}
\Delta E_\nu R_i=-\sum_j \bar f_{hh}(j,i)r_j+\sum_{j,a} \bar f_{hp}(j,a)r_{ji}^{a}%\\ \nonumber
-0.5\sum_{j,k,a} \bar V_{hhhp}(j,k,i,a)r_{kj}^{a} \,\,\, \forall \,\,i
\label{r1equation}
\end{eqnarray}
\begin{eqnarray}
\Delta E_\nu R^{a}_{ji}&=&
-\hat P(ij)\sum_k \bar f_{hh}(k,i)r_{ji}^{a}
-\sum_k \bar V_{hphh}(k,a,i,j)r_k+
\sum_{b} \bar f_{pp}(a,b)r_{ji}^{b}+
\hat P(ij)\sum_{k,c}\bar V_{phhp}(a,k,j,c)r_{ki}^{c} \nonumber \\
&+&\sum_{k,l}\bar V_{hhhh}(k,l,i,j)r_{lk}^c
+0.5\sum_{k,l,c,d} V_{hhpp}(k,l,c,d)r_{kl}^{c}t_{ij}^{db} \,\,\, \forall\,\,\, (a,\,\,j<i)
\label{r2equation}
\end{eqnarray}
\end{widetext}
Here $\bar f $, $\bar V$ and $V_{hhpp}(k,l,c,d)t_{ij}^{db}$  stands for one-body, two-body and three-body intermediate matrix elements.
The diagrams are evaluated by applying intermediate factorization scheme and a detail description can be found in Ref \cite{37}.
Since $\bar H$ is non-Hermitian in nature, therefore, there exists left eigenstates which are not identical to that of the right eigenstates but
they form a bi-orthogonal set, $\langle \Phi_0|L_\mu R_\nu|\Phi_0\rangle=\delta_{\mu \nu}$. These eigenstates can be obtained by 
diagonalizing $\bar {H}$ matrix. Both the left hand solution and the right hand solution leads to same eigenvalues. 
It is worth to note that for the calculation of energy, solution of any of the eigenstates is sufficient. In relativistic
calculation the dimension of the effective Hamiltonian matrix is huge. Therefore, it is not a good idea to follow a full 
diagonalization scheme.
We, therefore, implemented Davidson algorithm \cite{38} to diagonalize $\bar {H}$ matrix for the right hand solution. However, left eigenstates
are required along with the right eigenstates for the calculation of properties in EOMCC framework or to construct density matrix. 
Convergence is always of concern in the case of open-shell atomic and molecular systems. Therefore, Direct Inversion of Iterative Space (DIIS)
method is implemented to achieve a faster convergence for the solution of ground state amplitude equations. \par
%%%%%%%%%%%%%%%%%%%%%%%%%%%%%%%%%%%%%%%%%%%%%%%%%%%%%%%%%%%%%%%%%%%%%%%%%%%%%%%%%%%%%%%%%%%%%%%%%%%%%%%%%%%%%%%%%%%%%%%%%%%%%
In the case of EOM-MBPT(2), the ground state wavefunction is characterized at the MBPT(2) level, which require nothing but the
first order perturbed wavefunction. In this scheme the CCSD amplitudes are approximated at the MBPT(2) level. These amplitudes
can be written as 
\begin{eqnarray}
T_i^{a'}=\frac{f_{ia}}{\epsilon_i-\epsilon_a},\,\,\, 
T_{ij}^{ab'}=\frac{\langle ab||ij\rangle}{\epsilon_i+\epsilon_j-\epsilon_a-\epsilon_b}
\end{eqnarray}
with these non-iterative $T^\prime$ amplitudes one can calculate modified effective $\bar {H}$. 
This reduces computational scaling to a non-iterative $N^5$ instead of iterative $N^6$, where $N$ is the 
number of basis function. 
%%%%%%%%%%%%%%%%%%%%%%%%%%%%%%%%%%%%%%%%%%%%%%%%%%%%%%%%%%%%%%%%%%%%%%%%%%%%%%%%%%%%%%%%%%%%%%
%                        Basis Information
%%%%%%%%%%%%%%%%%%%%%%%%%%%%%%%%%%%%%%%%%%%%%%%%%%%%%%%%%%%%%%%%%%%%%%%%%%%%%%%%%%%%%%%%%%%%%%
\begin{table*}[t]
\caption{ Basis information and correlation energies from the MBPT(2) and CCSD method.}
\begin{ruledtabular}
{%
\newcommand{\mc}[3]{\multicolumn{#1}{#2}{#3}}
\begin{center}
\begin{tabular}{lrrrrrr}
Atom/ &\, Basis &\, Cutoff & \mc{2}{c}{Number of Spinor} & \mc{2}{c}{Correlation Energy}\\
\cline{4-5} \cline{6-7}
Molecule&\,  &\,(a.u.)&\,Occupied&\,Virtual&\,MBPT(2)&\, CCSD\\
\hline 
Ag&\,dyall.cv4z&\,$-$5000.0,\,500.0&\,47&\,487&\,$-$2.25310891&\,$-$2.14139607\\
Cs&\,dyall.cv4z&\,$-$5000.0,\,500.0&\,55&\,421&\,$-$1.97618962&\,$-$1.87252641 \\
Au&\,dyall.ae4z&\,$-$20.0,\,100.0&\,43&\,439&\,$-$2.43602044&\,$-$2.16853460\\
Fr&\,dyall.cv4z&\,$-$30.0,\,30.0&\,51 &\,363&\,$-$1.58857716&\,$-$1.44750991\\
Lr&\,dyall.cv4z&\,$-$50.0,\,20.0&\,67&\,519&\,$-$2.41494155&\, $-$2.04076559\\
HgH&\,Hg:dyall.cv3z&\,$-$25.0,\,1700.0&\,49&\,411&\,$-$2.61175077&\,$-$2.33935482\\
  ×&\,H:cc-pCVTZ&\,×&\,×&\,×&\,&\, \\
PbF&\,Pb:dyall.cv3z&\,$-$25.0,\,1700.0 &\,57&\,417&\,$-$1.99440949&\,$-$1.79400780\\
×&\,F:cc-pCVTZ&\,×&\,×&\,×&\,&\,
\end{tabular}
\end{center}
}%
\end{ruledtabular}
\label{tab1}
\end{table*}
%%%%%%%%%%%%%%%%%%%%%%%%%%%%%%%%%%%%%%%%%%%%%%%%%%%%%%%%%%%%%%%%%%%%%
\section{computational considerations}\label{sec3}
In this section, we address some computational aspects of the relativistic EOMCC method in the context of ionization potential. 
The spinor based EOMCC method as described in the previous section as listed in  Eqn \ref{r1equation} and Eqn \ref{r2equation}
has been coded as matrix multiplication, which results in a fully vectorized program for these ionization potential calculations.
We have generated one-body and two-body matrix elements required for the correlation calculations using DIRAC10 program package \cite{39}. 
Finite size nucleus is considered in all our calculations. The charge density of the finite size nucleus is defined using Gaussian
distribution model. All the parameters for the Gaussian distribution nuclear model is taken as default of DIRAC10 \cite{40}.
The scalar, real Gaussian functions are used to construct the finite atomic orbital basis. The wavefunction in the four-component relativistic
theory consists of both large and small component. The small component of the basis is generated by imposing restricted kinetic balance
condition (RKB). It generates both the large and small component of the wavefunction in equivalent amount. The employed RKB scheme properly represents the
kinetic energy (KE) solution in the non-relativistic limit and variational collapse is avoided which occurs due to the presence of both
electronic and positronic solution \cite{41}. As a results of which real finite eigen spectrum is achieved. In the process free particle Hamiltonian is
diagonalized to remove the unphysical solution. The uncontracted basis function better represents the relativistic wavefunction.
Therefore, we have taken the large component of the basis function in un-contracted fashion and small component is by default uncontracted
in DIRAC10.  
We adopted dyall.cv4z \cite{41a} basis for Ag and dyall.ae4z \cite{42} for Au in our calculations.
%We adopted dyall.ae4z \cite{42} basis for Ag and Au atom in our calculations.
In the calculations of Cs \cite{43}, Fr \cite{43}, and Lr \cite{44} atom, we have chosen
dyall.cv4z basis.
The high energy virtual orbitals contributes very less in a correlation calculation as energy difference between the virtual orbitals
and occupied orbitals appears in the denominator in coupled-cluster method. Therefore,
we have fixed a limit of 500 a.u. for the virtual orbital energies in the correlation calculation of Ag and Cs 
atom. None of the core electrons are frozen for these two atoms.  
For Au, Fr, and Lr we have frozen 36 inner core electrons and a cutoff of 100 a.u., 30 a.u. and 20 a.u. is fixed for the virtual orbitals, respectively. 
Besides these atomic systems we have calculated ionization potentials of PbF and HgH. We have chosen dyall.cv3z basis for the Hg \cite{44a} 
atom and cc-pCVTZ basis for the H atom \cite{45}.  
Dyall.cv3z basis is opted for the Pb atom \cite{46} and cc-pCVTZ basis for F \cite{45}.
All the orbitals in between -25 a.u. and 1700 a.u. is taken into consideration for the correlation calculation for both the molecular system.
%and 36 core electrons are frozen for PbF molecule and cutoff of 1700 a.u. is fixed for the virtual orbital energy. 
We have taken experimental bond length 2.0585{\AA} for PbF and 1.7660 {\AA} for HgH molecule from Ref. \cite{47}. 
A cutoff $10^{-12}$ is fixed while storing the two-body matrix elements required for the ground state energy calculation as the contribution of 
the two-body matrix elements beyond 12 digit after the decimal point is negligible for the correlation calculation.
%A convergence of $10^{-9}$
%is fixed for the solution of coupled cluster amplitude equation except for the Lr atom where we have used $10^{-7}$ as convergence
% cut off and $10^{-5}$ in the EOM part for all the calculations.
%A DIIS space of 6 is used for the solution of ground state amplitude equations and we have changed it 8 for Lr atom after reaching 
%convergence of $0.37887830\times 10^{-6}$ due to its very slow convergence pattern.  
A convergence cutoff of $10^{-7}$ is fixed for the norm of the error vector in the SCF calculation except for Lr atom where a cutoff 
of $10^{-5}$ is used. For ground state coupled-cluster amplitude calculation, we have used a DIIS space of 6 and $10^{-9}$
as a convergence cutoff except for Lr. In case of Lr, we have used a DIIS space of 6 upto the convergence of $0.37\times 10^{-6}$
and then a DIIS space of 8 is used until it reaches to $10^{-7}$. The convergence cutoff used for all the EOMCC calculation is $10^{-5}$. 
\section{results and Discussion}\label{sec4}
%%%%%%%%%%%%%%%%%%%%%%%%%%%%%%%%%%%%%%%%%%%%%%%%%%%%%%%%%%%%%%%%%%%%%%%%%%%%%%%%%%%%%%%%%%%%%%
%                         IP Values
%%%%%%%%%%%%%%%%%%%%%%%%%%%%%%%%%%%%%%%%%%%%%%%%%%%%%%%%%%%%%%%%%%%%%%%%%%%%%%%%%%%%%%%%%%%%%%
\begin{table*}[ht]
\caption{ Experimental and theoretical IP values (in eV) in MBPT(2), RPA and CCSD approximation}
\begin{ruledtabular}
\begin{center}
\begin{tabular}{llrrrrr}
Atom/ &\,Ionized &\,MBPT(2)&\,RPA&\,CCSD&\,Others&\,Experiment\\ %%%%%%%%%%%%%%%%%%%%%%%%%%%%%%%%%%%%%%%%%%%%%%%%%%%%%%%%%%%%%%%%%%%%%%%%%%%%%%%%%%%%%%%
Molecule       &\,state   &\,       &\,   &\,     &\,     &\,          \\                      %%%%%%%%%%%%%%%%%%%%%%%%%%%%%%%%%%%%%%%%%%%%%%%%%%%%%%%%%%%%%%%%%%
\hline                                                                                                     %%%%%%%%%%%%%%%%%%%%%%%%%%%%%%%%%%%%%%%%%%%%%%
Ag &\,$^1S$ &\,7.5155 &\,7.8183 &\,7.5118 &\,7.5411\cite{48},\,7.2368\cite{49}&\,7.5762\cite{51}\\             %%%%%%%%%%%%%%%%%%%%%%%%%%%%%%%%%%%%%%%%%%
   &\,      &\,        &\,        &\,     &\,7.3600\cite{50},\,7.4610\cite{50}&\,                 \\               %%%%%%%%%%%%%%%%%%%%%%%%%%%%%%%%%%%%%%
   &\,$^1D$ &\, 12.9632&\, 16.6037 &\, 12.4269&\,                               &\,                 \\                %%%%%%%%%%%%%%%%%%%%%%%%%%%%%%%%%%%%    
   &\,$^3D_1$&\,12.9958 &\, 16.6303&\,12.4605&\,                               &\,                 \\              %%%%%%%%%%%%%%%%%%%%%%%%%%%%%%%%%%%%%%   
   &\,$^3D_2$&\,13.0321 &\,16.6586 &\, 12.4974&\,                              &\,                 \\             %%%%%%%%%%%%%%%%%%%%%%%%%%%%%%%%%%%%%%% 
   &\,$^3D_3$&\,13.0719&\,16.6892  &\, 12.5388&\,                              &\,                 \\           %%%%%%%%%%%%%%%%%%%%%%%%%%%%%%%%%%%%%%%%%
%%%%%%%%%%%%%%%%%%%%%%%%%%%%%%%%%%%%%%%%%%%%%%%%%%%%%%%%%%%%%%%%%%%%%%%%%%%%%%%%%%%%%%%%%%%%%%%%%%%%%%%%%%%%%%%%%%%%%%%%%%%%%%%%%%%%             %%%%%%%%
  \\                                                                                                                                   %%%%%%%%%%%%%%%%%
%%%%%%%%%%%%%%%%%%%%%%%%%%%%%%%%%%%%%%%%%%%%%%%%%%                                                  %%%%%%%%%%%%%%%%%%%%%%%%%%%%%%%%%%%%%%%%%%%%%%%%%%%%%                    
Cs &\,$^1S$ &\, 3.8978 &\,3.9585&\,3.9016&\,3.8984\cite{52},\,3.8983\cite{52},\,3.7497 \cite{52}&\, 3.8939\cite{51} \\                     %%%%%%%%%%%%%%
   &\,      &\,        &\,      &\,      &\,3.8795\cite{53,54}                                  &\,                  \\                  %%%%%%%%%%%%%%%%
   &\,$^1P$ &\,17.9767 &\,19.7254&\,17.5515&\,                                                  &\,                 \\              %%%%%%%%%%%%%%%%%%%%%
   &\,$^3P_0$&\,18.0156&\,19.7462&\,17.5911&\,                                                  &\,                  \\           %%%%%%%%%%%%%%%%%%%%%%%
   &\,$^3P_1$&\,18.0576&\,19.7674&\,17.6338&\,                                                  &\,                   \\         %%%%%%%%%%%%%%%%%%%%%%%%
   &\,$^3P_2$&\,18.1041&\,19.7890&\,17.6807&\,                                                  &\,                    \\              %%%%%%%%%%%%%%%%%%
%%%%%%%%%%%%%%%%%%%%%%%%%%%%%%%%%%%%%%%%%%%%%%%%%%%%%%%%%%%%%%%%%%%%%%%%%%%%%%%%%%%%%%%%%%%%%%%%%%%%%%                       %%%%%%%%%%%%%%%%%%%%%%%%%%%%
 \\                                                                                                          %%%%%%%%%%%%%%%%%%%%%%%%%%%%%%%%%%%%%%%%%%%
%%%%%%%%%%%%%%%%%%%%%%%%%%%%%%%%%%%%%%%%%%%%%%%%%%%%%%%%%%%%%%%%%%%%%%%%%%%%%%%%%%%%%%%%%%%%%%%%%%%%%%%%%%%%%%                               %%%%%%%%%%%%   
%%%%%%%%%%%%%%%%%%%%%%%%%%%%%%%%%%%%%%%%%%%%%%                                                                      %%%%%%%%%%%%%%%%%%%%%%%%%%%%%%%%%%%%%
Au &\,$^1S$ &\,9.3467&\,9.7898&\,9.2103&\, 9.1010\cite{55},\,9.0860\cite{55},\,6.9810\cite{55}&\,9.2255 \cite{51}\\                       %%%%%%%%%%%%%%%
   &\,      &\,      &\,      &\,      &\, 9.0340\cite{50},\,9.1230\cite{50}                  &\,                \\                     %%%%%%%%%%%%%%%%%
   &\,$^1D$ &\,11.5241&\,14.3886&\,10.9553&\,                                &\,                  \\                                   %%%%%%%%%%%%%%%%%%
   &\,$^3D_1$&\,11.5776&\,14.4447&\,11.0109&                                 & \,                  \\                                       %%%%%%%%%%%%%
   &\,$^3D_2$&\,11.6340&\,14.5037&\,11.0703&\,                               &\,                    \\                                   %%%%%%%%%%%%%%%%
   &\,$^3D_3$&\,11.6949&\,14.5660&\,11.1342&\,                               &\,                     \\                               %%%%%%%%%%%%%%%%%%%
%%%%%%%%%%%%%%%%%%%%%%%%%%%%%%%%%%%%%%%%%%%%%%%%%%%%%%%%%%%%%%%%%%%%%%%%%%%%%%%%%%%%%%%%%%%%%%%%%%%%                             %%%%%%%%%%%%%%%%%%%%%%%%
 \\                                                                                               %%%%%%%%%%%%%%%%%%%%%%%%%%%%%%%%%%%%%%%%%%%%%%%%%%%%%
%%%%%%%%%%%%%%%%%%%%%%%%%%%%%%%%%%%%%%%%%%%%%%%%%%%%%%%%%%%%%%%%%%%%%%%%%%%%%%%%%%%%%%%%%%%%%%%%%%%%%%%%%%%%%                                   %%%%%%%%%    
Fr &\,$^1S$ &\,4.0890 &\,4.1635 &\,4.0790 &\,4.0715\cite{52},\,3.6041 \cite{52}  &\,4.0727\cite{51} \\                                          %%%%%%%%%
   &\,$^1P$ &\,15.5869 &\,17.0747&\,15.1745&\,                            &\,                   \\                                               %%%%%%%%
   &\,$^3P_0$&\,15.6346&\,17.1084&\,15.2233&\,                           &\,                  \\                                                  %%%%%%%
   &\,$^3P_1$ &\,15.6880&\,17.1423 &\,15.2762  &\,                           &\,                 \\                                       %%%%%%%%%%%%%%%
   &\,$^3P_2$ &\, 15.7455 &\,17.1764 &\, 15.3340&\,                          &\,                   \\                                     %%%%%%%%%%%%%%%
%%%%%%%%%%%%%%%%%%%%%%%%%%%%%%%%%%%%%%%%%%%%%%%%%%%%%%%%%%%%%%%%%%%%%%%%%%%%%%%%%%%%%%%%%%%%%%%%%%%%%%%%%%%%%%%%%                        %%%%%%%%%%%%%%%%
  \\                                                                                                                                   %%%%%%%%%%%%%%%%%%
%%%%%%%%%%%%%%%%%%%%%%%%%%%%%%%%%%%%%%%%%%%%%%%%%%%%%%%%%%%%%%%%%%%%%%%%%%%%%%%%%%%%%%%%%%%%%%%%%%%%%%%%%%%%%%%%%                                  %%%%%%
%%%%%%%%%%%%%%%%%%%%%%%%%%%%%%%%%%%%%%%%%%%%%%%%                                                                                               %%%%%%%%%%
Lr &\,$^1S$ &\, 5.3353 &\, 6.0794  &\, 4.8715&\,4.8870\cite{57},\,4.8931\cite{58},\,4.8936\cite{58} &\,4.96$^{+0.08}_{-0.07}$\cite{60} \\      %%%%%%%%%%
   &\,      &\,        &\,         &\,       &\,4.9340\cite{59},\,4.963(15)\cite{60}    &\,                              \\                   %%%%%%%%%%%  
   &\,$^1P$ &\,7.9047  &\,8.0777  &\, 7.5281 &\,                                           &\,                              \\               %%%%%%%%%%%%
   &\,$^3P$ &\, 8.0437 &\,8.2979  &\, 7.7090 &\,                                           &\,                               \\              %%%%%%%%%%%%
%%%%%%%%%%%%%%%%%%%%%%%%%%%%%%%%%%%%%%%%%%%%%%%%%%%%%%%%%%%%%%%%%%%%%%%%%%%%%%%%%%%%%%%%%%%%%%%%%%%%%%%%%%%%%%%%%                          %%%%%%%%%%%%%%
  \\                                                                                                                                      %%%%%%%%%%%%%%%
%%%%%%%%%%%%%%%%%%%%%%%%%%%%%%%%%%%%%%%%%%%%%%%%%%%%%%%%%%%%%%%%%%%%%%%%%%%%%%%%%%%%%%%%%%%%%%%%%%%%%%%%%%%%%%%%%%                        %%%%%%%%%%%%%%%  
%%%%%%%%%%%%%%%%%%%%%%%%%%%%%%%%%%%%%%%%%%%%%%%%%%%%%%%%                                                                                   %%%%%%%%%%%%%%  
HgH &\,$^1\Sigma$ &\, 8.2228 &\, 9.2480 &\, 8.2319&\,7.8350\cite{61},\,7.8660\cite{61},\,7.9600\cite{61}        &\,8.60 $\pm$0.43\cite{62}\\%%%%%%%%%%%%%
%%%%%%%%%%%%%%%%%%%%%%%%%%%%%%%%%%%%%%%%%%%%%%%%%%%%%%%%%%%%%%%%%%%%%%%%%%%%%%%%%%%%%%%%%%%%%%%%%%%%%%%%%%%%%%%%%%%%%%%%%%%%%%%%%%%%%%%%%%%%%%%%%%%%%%%%%
\\                                                                                                                                          %%%%%%%%%%%%%
%%%%%%%%%%%%%%%%%%%%%%%%%%%%%%%%%%%%%%%%%%%%%%%%%%%%%%%%%%%%%%%%%%%%%%%%%%%%%%%%%%%%%%%%%%%%%%%%%%%%%%%%%%%%%%%%%%%%%%%%%%%%%%%%%%%%%%%%%%%%%%%%%%%%%%%%%
PbF&\,$^1\Sigma$ &\,7.7491&\,8.6091&\,7.5464&\,6.7756\cite{63}                     &\,7.54(1)\cite{64}                                %%%%%%%%%%%%%%%%%%%       
%%%%%%%%%%%%%%%%%%%%%%%%%%%%%%%%%%%%%%%%%%%%%%%%%%%%%%%%%%%%%%%%%%%%%%%%%%%%%%%%%%%%%%%%%%%%%%%%%%%%%%%%%%%%%%%%%%%%%%%%%%%%%%%%%%%%%%%%%%%%%%%%%%%%%%%%%
\end{tabular}
\end{center}
\end{ruledtabular}
\label{tab2}
\end{table*}
%%%%%%%%%%%%%%%%%%%%%%%%%%%%%%%%%%%%%%%%%%%%%%%%%%%%%%%%%%%%%%%%%%%%%%%%%%%%%%%%%%%%%%%%%%%%%%
The numerical results of our calculation are presented to judge the accuracy obtained in our newly 
implemented open-shell reference relativistic EOMCC method for the single electron ionization problem. We have chosen heavy atomic (Ag, Cs, Au, Fr, Lr) and 
molecular (HgH, PbF) systems where both the effect of electron correlation and relativistic effects are important
to obtain high accurate results.
We have calculated a few of the low-lying both singlet and triplet ionized states with their fine structure splitted states for the atomic systems 
and for the molecular systems, only the valence ionization potential is reported.  
Three different sets of calculation are done for all the considered atomic 
and molecular systems, viz EOMCC using CCSD ground state wavefunction as reference, MBPT(2)
ground state wavefunction as reference and RPA calculation on top of CCSD ground state wavefunction.
All these calculated results 
are compared with the experimental values and with the calculations based on other theoretical means. The intermediate
calculation in the four-component EOMCC framework will help us to understand how the effects of correlation play role to get good 
agreement with the sophisticated experiments. \par 
%%%%%%%%%%%%%%%%%%%%%%%%%%%%%%%%%%%%%%%%%%%%%%%%%%%%%%%%%%%%%%%%%%%%%%%%%%%%%%%%%%%%%%%%%%%%%%%%%%%%%%%%%%%%%%%%%%%%%%%%%%%%%%%%%%%%%
In Table \ref{tab1}, we have provided information regarding the basis set employed, cutoff used in the orbital energies for the
correlation calculation and the dimension of the determinantal space constructed from number of occupied and virtual orbitals
used in the correlation calculation for all the considered atomic and molecular systems. Further, we have reported the correlation
energies from both second-order perturbation theory level (MBPT(2)) and using coupled-cluster single- and double- excitation
approximation (CCSD) method. It is worth to note that CCSD correlation energy from our calculation is matching up to 7-digit
after the decimal place with DIRAC10 irrespective of the choice basis set and the system of study. The discrepancy beyond this limit
could be due to the use of different convergence algorithm and the use of cutoff in storing two-body matrix elements. \par
%%%%%%%%%%%%%%%%%%%%%%%%%%%%%%%%%%%%%%%%%%%%%%%%%%%%%%%%%%%%%%%%%%%%%%%%%%%%%%%%%%%%%%%%%%%%%%%%%%%%%%%%%%%%%%%%%%%%%%%%%%%%%%%%%%%%%%%%%%
The calculated ionization potential values of the considered atomic and molecular systems using different approximation scheme
in the EOMCC framework are reported in Table \ref{tab2}. A comparison of these calculated values is made with the values 
calculated using other theoretical methods. Finally all these values are compared with the experimental values to judge 
how different many-body effects play role in getting accurate results in the considered systems. \par
\subsection{Ag atom}
Ag is a moderate heavy atomic system. Therefore, relativistic treatment is necessary to obtain accurate result.
There is a single electron outside of the closed-shell core and the ground state configuration of Ag is [Kr]4$d^{10}5s^{1}$.
The ionization potential calculated using EOMCC method with CCSD as the ground state reference wavefunction
is 7.5118 eV. The MBPT(2) EOMCC results are very close with the
CCSD results. Therefore, missing dynamic correlation does not have much effect in the calculated ionization potential
values of the Ag atom. On the other hand, RPA result is much higher than the CCSD results, since in the RPA scheme,
construction of the EOM matrix elements are done only in the 1-h excited determinantal space, therefore, a major part of the non-dynamic
electron correlation is missing. Thus, it can be said that the non-dynamic part of the electron correlation is more 
important than the dynamic part for Ag atom. Neogrady {\it et al.} \cite{50} have done ionization potential calculation
of Ag atom with $4s^{2}4p^{6}4d^{10} 5s^{1}$ shell is used for the  
correlation calculation and % whereas we have taken into account of all the electrons in correlation calculation.
spin free Douglas-Kroll (D-K) Hamiltonian is used for the purpose of relativistic treatment. 
Two different sets of calculation are done; one with CCSD ground state wavefunction and other with the 
further addition of partial triples. There is a difference 
of 0.1 eV. in these two employed schemes. Therefore, it can be said that consideration of partial
triples is important which is evident from Neogrady {\it et al.}'s calculation, however, it can not be quantified since different many-body effects are 
intertwined. Safronova {\it et al.} \cite{49} also calculated ionization potential of 
Ag atom at the level of third order perturbation theory (MBPT(3)) with the consideration
of relativistic effect using Dirac-Coulomb-Breit Hamiltonian and the correction due to the Lamb shift is also included. 
Therefore, the relativistic treatment is far more comprehensive than any other approach but 
the calculated result is poor in comparison to any other calculation as the effect 
of electron correlation is not considered using a high level of correlated theory. Nayak {\it et al.} \cite{48} 
used valence-universal or the Fock space multi-reference coupled cluster theory in their
calculation using closed-shell reference CCSD ground state wavefunction. Their result is 0.03 eV better 
than our calculated result. They have used 36s30p28d15f primitive GTOs in their calculations whereas
we have used dyall.cv4z basis which uses 33s25p17d9f6g3h primitive GTOs. 
Nayak {\it et al.} did not use any cutoff for the virtual orbital energies in their
calculation but considered Fermi distribution nuclear model to take care of the finite 
size of the nucleus and exploited spherical symmetry. It is worth note that the Fermi distribution nuclear model
is a better representation of an atomic nucleus than the Gaussian distribution. However, for molecular codes, consideration of Fermi distribution nuclear model 
is not favorable from the computational perspective. 
The spherical implementation allows the separation of the radial and angular part to use the reduced matrix elements.
The calculations based on reduced matrix elements are easy from computational point of view as it only requires evaluation
of the radial integrals and the angular part is add up to as a multiplier.
The reason for the difference between our CCSD and the NIST data 
could possible be due to the dominance of double-excitation character in Ag atom.
The EOMCC method with single- and double- excitation approximation accounts contribution of 
single excitation and to some extent double excitations.
However, this could be attenuated by the inclusion of the triple excitations both in the ground
and excited states.
The difference between our calculated ionization potential value using CCSD reference wavefunction with NIST
database is about 0.06 eV which is in the error limit of 0.9 \%. \par
\subsection{Cs atom}
The alkali metal, Cs has the ground state electronic configuration $[Xe]6s^1$. The effect of relativity
will not be that much as compared to Ag atom as there is a single electron outside of a noble gas core, whereas
there are ten d-electrons outside of the Kr core in Ag atom. The effect of relativity has much larger influence
on the d orbitals than on p orbitals.
%The d orbitals are get more effected than the p orbitals due to effect 
We have done three sets of calculation for
Cs atom and compared with the values calculated by using other theoretical methods and with experimental values. 
Eliav {\it et al.} \cite{52} have done three different sets of calculation for Cs atom. They have used both
Dirac-Coulomb and Dirac-Coulomb-Breit Hamiltonian in their calculations and one calculation at the
non-relativistic limit taking speed of light as $10^5$ and the correlation treatment is done using FSMRCC theory starting from a closed-shell
CCSD reference wavefunction.
The results calculated using DC Hamiltonian
and DCB Hamiltonian is 0.0001 eV. 
Therefore, the result of their calculation justifies that higher order relativistic effect 
specially the Breit interaction does not have much effect in the calculated ionization potential value of Cs atom. The difference
between non-relativistic and relativistic calculation is only about 0.15 eV.
They followed a strategy
like $M^{+}(0,0) \rightarrow  M(0,1) \rightarrow M^{-}(0,2)$ to calculate ionization potential, excitation energy, and electron affinity 
value of the open-shell system using effective Hamiltonian based relativistic FSMRCC method. Therefore, such an approach is not healthy from the
computational point of view. They have correlated ns,(n-1)s,(n-1)p, and (n-1)d orbitals and a cutoff
of 100-150 a.u. for the virtual orbital energies whereas we have used 500 a.u. as cutoff and correlated all the electron.
Our results calculated
using EOMCC method with CCSD reference wavefunction and the Eliav's results are almost similar.
Therefore, deep core orbitals
of the alkali metals can be frozen for the correlation calculation as these electrons does not have much influence in the calculated ionization potential values. 
If we look into our calculated results the MBPT(2) results are very close to the CCSD results than RPA
results but the effect of non-dynamic correlation is not that much what we observed in the case of Ag atom
as there is only 0.05 eV. difference between CCSD and RPA results. It can be said that MBPT(2) is better
approximation than the RPA for the case of Cs atom also. 
Blundell and coworker \cite{53,54} also
done ionization potential calculation using a linear version of relativistic CC theory considering
single-, double-, and an important subset of the triple excitation. The difference between their result
and the NIST value is about 0.15 eV. Therefore, it can be concluded that non-linear terms have 
significant contribution to the calculated ionization potential values. \par
\subsection{Au atom}
The relativistic effect of Au is a well known phenomenon. The non-relativistic calculation of Au 
shows incorrect numbers, even it results in the reversal in the order of first two excited states. 
Therefore, a relativistic description is must required for calculation of ionization potential of 
Au atom. The electronic configuration of Au in the ground state is $[Xe]4f^{14}5d^{10}6s^{1}$.   
We have done three sets of calculation for the Au atom as previously mentioned. 
All these results are compared with the results calculated by Eliav and coworkers \cite{55} and  
Neogrady {\it et al.} \cite{50}. Neogrady {\it et al.}. employed spin-free D-K Hamiltonian in their
calculations and there is a difference about 0.08 eV. in CCSD and CCSD(T) calculations. Therefore,
it can said that triple excitation have significant contribution to the calculated ionization
potential value of Au atom. They have considered only the $5p^65d^{10}6s^1$ shell for beyond SCF
calculation. If we look at the Eliav {\it et al}'s results, the non-relativistic result
is totally out of track. There is a difference of about 2 eV. with the experiment.
The difference between calculation based on DC Hamiltonian and DCB Hamiltonian is about 
0.02 eV, therefore higher order relativistic effect have role in the calculated 
values. Eliav {\it et al.} used same strategy in the FSMRCC framework as mentioned in the
discussion of Cs atom  and used a rather small basis consists of only 21s17p11d7f GTOs.
If we look into our calculated results, it can be observed that 
%Our EOMCC results are much more accurate than any other calculation.
the non-dynamic part of the electron correlation has larger influence than the dynamic part as the difference
between MBPT(2) result and EOMCC is about 0.13 eV whereas it is 0.6 eV for CCSD and RPA. 
We have achieved a better accuracy using our open-shell reference EOMCC method than
the calculation based on FSMRCC by Eliav and coworkers. 
The error in our calculation is 0.1 \% in comparison to the experimental value. \par
\subsection{Fr atom}
Fr is heavier than the Au atom but effect of relativity is not that much 
as the difference between relativistic results and non-relativistic result
is about 0.5 eV. in comparison to 2 eV. in the case of Au atom which is obtained from Eliav {\it et al.} \cite{52} calculations. They 
employed effective Hamiltonian formalism of the FSMRCC theory to calculate ionization potential values of Fr atom
starting from a closed-shell reference state wavefunction. From the 
reported result of Cs, it is evident that higher order relativistic effects
specially the Breit interaction does not have much effect in the case of 
alkali metal atoms.  Eliav {\it et al.} used well tempered s-basis consists of
36s26p21d15f10g primitive GTOs for the Fr atom.
The atomic orbitals having same l value but with different k quantum numbers
are expanded using same set of basis functions. The outer orbitals and 
(n-1)s,(n-1)p, (n-1)d orbitals are correlated and a cutoff 100-150 a.u. is used. 
%Dzuba {\it el al.} \cite{56}
The calculated EOMCC results are almost identical to that of FSMRCC results.
In our EOMCC calculations it is observed that the difference between MBPT(2)
and CCSD results is about 0.01 eV. On the other hand the difference is 0.07 eV
for RPA and CCSD. The calculated result justifies that the missing 2h-1p block in the RPA
scheme does not have much influence in the calculated ionization potential value
of Fr in comparison to what we observed in Ag and Au atom. The accuracy achieved
in our EOMCC calculation is 0.02 \%. \par
\subsection{Lr atom}
The effect of relativity play a decisive role in the determination of electronic structure of the heavy elements
in the sixth row of the periodic table. This effect is profoundly increases in 
the seventh row including the actinides. The effect is such that it even alter
the ground state electronic configuration. The $s_{1/2}$ and $p_{1/2}$ are stabilized
whereas $p_{3/2}$, $d_{5/2}$, $d_{7/2}$ and $f_{7/2}$, $f_{9/2}$ orbitals are destabilized 
which makes their ground state configuration different from the lighter elements of the 
same group. The ground state electronic configuration of Lr was debatable for years.
A recent successful measurement and theoretical calculation confirmed $[Rn]5f^{14}7s^27p^1$ ground state configuration
in contrast to $[Xe]4f^{14}6s^25d$ of its homologue Lu. 
There are various theoretically calculated value available for the ionization potential of Lr atom using
different variants of relativistic correlated theories.
Eliav {\it et al.} \cite{57} applied effective Hamiltonian variant of the FSMRCC theory to calculate ionization
potential value of Lr starting from a closed-shell reference ground state wavefunction and the
relativistic treatment is done using DCB Hamiltonian. They have correlated 5df6spd7sp shell
and rest of the inner electrons are frozen. A cutoff of 100 a.u. is used for the virtual orbital energies and they
got 4.8870 eV. as ionization potential value of Lr. 
Borschevsky {\it et al.} \cite{58} applied intermediate Hamiltonian variant of the FSMRCC theory using a larger model 
space with DCB Hamiltonian. The inclusion of QED effect does not have much effect in the calculated 
ionization potential value as there is a change of 0.0005 eV. with inclusion of the QED effects.
The QED effects are important for the highly charged ion rather for neutral or mono-cationic ions. 
They used universal basis set with even-tempered Gaussian type of orbitals (37s, 31p, 26d, 21f, 16g, 11h, and 6i
orbitals) in uncontracted fashion with a cutoff of 200 a.u. for the virtual orbital energies and 
correlated only 43 outer electrons in their calculation and got 4.8931 eV. as the valence
ionization potential value. 
Dzuba {\it et al.} \cite{59} applied CI+all order method to calculate ionization potential of Lr atom. 
In this approach ground state wavefunction is constructed using linearized version of the
CCSD method. The correlation between valence electrons are treated with CI while
core-valence electron correlation is taken care by the linearized version of CCSD method.
The relativistic effect is taken care using DCB Hamiltonian with inclusion of QED effect and 4.9340 eV.
is obtained as the ionization potential value. The approach can be regarded as the Hilbert space analogue
of the EOMCC method which is a Fock space approach.
If we look into our results, the difference between EOMCC result and that of the latest experiment 
is about 0.09 eV. This difference could be reduced with the inclusion of larger correction space as
we have only taken 20 a.u. as virtual orbital energy cutoff, missing Breit interaction, triple excitation or the
finite size of the basis. However, it is difficult to blame any of the effect as all these effects are intricately
coupled to each other. There is a difference of more than 1 eV. between CCSD and RPA values but MBPT(2) result
is more accurate than RPA though effect is quite high. Therefore,
it can be inferred that both dynamic and non-dynamic correlation effect have greater role in Lr.
The best result in accordance to the measured value
is obtained using delta CCSD(T) method with DC Hamiltonian plus the Breit correction and Lamb Shift \cite{60}.
The Breit correction (-12 meV) and the Lamb shift (16 meV) does not have much effect on the calculated IP value.  
Two different calculation are done, one for the neutral atom and one for the mono-cationic ion and the
difference is reported as the IP value of Lr atom. 
The EOMCC approach due to its CI like structure is capable of taking care of non-dynamic electron correction elegantly and performs
better in comparison to $\Delta$CC approach for the treatment of non-dynamic electron correlation.
However, the treatment of dynamic correlation is inferior with CCSD ground state wavefunction
since a complete treatment of doubly excited configuration relative to the reference determinant require at least a part of the $T_3$ operator.
Lr is very heavy atomic system, therefore there is a huge contribution of the relaxation effect in the computed ionization potential value 
and the $\Delta$ based methods are more elegant in taking account of the relaxation effect than any energy difference method. \par
\subsection{HgH}
The results for HgH using all three variant of the relativistic EOMCC method using open-shell reference wavefunction
is compiled in Table \ref{tab2}. We have compared our calculated results with the values from experiment and other theoretically
calculated values. The outcome of our calculation shows that MBPT(2) results are very close to CCSD result, on the other hand, 
the difference between RPA and CCSD result is about 1 eV.
Our calculated CCSD results are in best agreement with the experiment though
there is a huge uncertainty in the measured value. 
H\"{a}ussermann {\it et al} \cite{61} calculated IP of HgH using quasi-relativistic pseudo-potential method with correlation treatment
is done using CISD, MRCI and CASSCF+MRCI with the addition of QED effects. It is true that there is not much effect of QED
in neutral molecule. The result of their calculation lacks both proper relativistic treatment and electron correction. As 
a result of which the calculated results in all three approach are not in accordance with the experimental value. \par
\subsection{PbF}
PbF has the ground state term $^{2}\Pi_{\frac{1}{2}}$, that means the unpaired electron is in the $\pi$ orbital.
We have calculated valence IP value of PbF using different approximation scheme in the relativistic EOMCC
framework starting from a open-shell reference wavefunction and presented the result in Table \ref{tab2}
along with other atomic or molecular systems. We have compared our results with 
Kozlov {\it et al.}'s \cite{62} result and with the experimentally reported value.
Our CCSD result is found to be more accurate than them. The MBPT(2) result is 
very close to the CCSD result whereas RPA results are very poor. Therefore,
it can be inferred that non-dynamic electron correlation is dominant in PbF than that
of dynamic part. Our CCSD result is in the accuracy of less than 0.1 \%. \par
%This could further be 
%improved with the inclusion of triple excitation in the construction of ground state
%wavefunction. 
%%%%%%%%%%%%%%%%%%%%%%%%%%%%%%%%%%%%%%%%%%%%%%%%%%%%%%%%%%%%%%%%%%%%%%%%%%%%%%%%%%%
%\clearpage
%\newpage
\section{conclusion}\label{sec5}
We have successfully implemented relativistic equation-of-motion coupled-cluster method using open-shell reference wavefunction.
%to calculate ionization potential values of heavy atomic and molecular systems in their open-shell ground state configuration.
The implemented method is employed to calculate a few of the low-lying ionized states of heavy atomic systems (Ag, Cs, Au, Fr, Lr) as well as 
valence ionization potential of HgH and PbF. The calculated results are compared with the experimental values as well
as other theoretical values and it is observed that our calculated results are very accurate.
%The calculated MBPT(2) results are much closer to the CCSD results than RPA results. Therefore,
%it can be inferred that non-dynamic correlation have greater effects than the missing
%dynamic part in the computed ionization potential values of the heavier systems.
\section*{acknowledgments}
H.P., S.S., and N.V. acknowledge the grant from CSIR XIIth five year plan project on Multi-scale Simulations of 
Material (MSM) and facilities of the Center of Excellence in Scientific Computing at CSIR-NCL.
H.P and S.S acknowledge the 
Council of Scientific and Industrial Research (CSIR) for their fellowship. S.P. acknowledges grant from DST, 
J. C. Bose fellowship project towards completion of the work.
%\clearpage
%\newpage

\end{document}